\input harvmac
\input epsf
\def\figin{\epsfcheck\figin}\def\figins{\epsfcheck\figins}
\def\epsfcheck{\ifx\epsfbox\UnDeFiNeD
\message{(NO epsf.tex, FIGURES WILL BE IGNORED)}
\gdef\figin##1{\vskip2in}\gdef\figins##1{\hskip.5in}
\else\message{(FIGURES WILL BE INCLUDED)}%
\gdef\figin##1{##1}\gdef\figins##1{##1}\fi}
\def\DefWarn#1{}
\def\figinsert{\goodbreak\midinsert}
\def\ifig#1#2#3{\DefWarn#1\xdef#1{fig.~\the\figno}
\writedef{#1\leftbracket fig.\noexpand~\the\figno}%
\figinsert\figin{\centerline{#3}}\medskip\centerline{\vbox{\baselineskip12pt
\advance\hsize by -1truein\noindent\footnotefont{\bf Fig.~\the\figno:} #2}}
\bigskip\endinsert\global\advance\figno by1}

\def\npb{Nucl.Phys.}
\def\prd{Phys. Rev. }
\def\plb{Phys. Lett. }
\def\bb#1{ hep-th/#1}
\def\dd{\partial }
\def\ovk{1\over k}

\lref\vafa{C. Vafa, {\it Gas of D-Branes and Hagedorn Density of BPS States},
hep-th/9511088, Nucl.Phys. B463 (1996) 415-419.}

\lref\braneReview{D. Kutasov and E. Giveon, Rev.Mod.Phys. 71 (1999) 983-1084
, hep-th/9802067}
\lref\cw{S. Coleman and  E. Weinberg, Phys.Rev.D7:1888-1910,1973}
\lref\books{Books on string theory such as M. Green, J. Schwarz, and  E.
Witten
Cambridge, Uk: Univ. Pr. ( 1987) and
J. Polchinski, Cambridge, Uk: Univ. Pr. (1998)}
\lref\wb{J. Wess and J. Baggar, Princeton University Press 1992}
\lref\bss{T. Banks, N. Seiberg, and E. Silverstein,
\plb B401:30-37,1997; \bb9703052.}
\lref\bdg{C. Bachas, M. Douglas, and M. Green, JHEP 9707:002, 1997;
\bb9705074.}
\lref\kMc{S. Kachru and J. McGreevy, \prd D61:026001,2000; \bb9908135.}
\lref\bdl{M. Berkooz, M. Douglas, and R. Leigh, \npb B480:265-278, 1996;
\bb9606139.}
\lref\dkps{M. Douglas, D. Kabat, P. Pouliot, and S. Shenker
\npb B485:85-127,1997; \bb9608024.}
\lref\az{A. Hanany and A. Zaffaroni, \npb B509:145-168, 1998;
\bb9706047.}

\lref\bh{
J.~H.~Brodie and A.~Hanany,
``Type IIA superstrings, chiral symmetry, and N = 1 4D gauge theory  dualities,''
Nucl.\ Phys.\  {\bf B506}, 157 (1997)
[hep-th/9704043].}

\lref\romans{L.J. Romans, \plb B169:374, 1986.}
\lref\ght{M. Green, C. Hull, and P. Townsend,
\plb 382(1996)65-72;\bb9604119.}
\lref\strominger{A. Strominger, \plb B383:44-47, 1996; \bb9512059.}
\lref\ch{C.G.Callan and J.A. Harvey, \npb B250:427,1985.}
\lref\brodie{J.H.Brodie, \npb B532:137-152,1998; \bb9803140.}
\lref\CSreview{For a nice review see
G.V. Dunne, Les Houches Lectures 1998; \bb9902115.}
\lref\jl{J. Maldacena and L. Susskind, \npb B475:679-690,1996;
\bb9604042.}
\lref\redlich{Redlich, \prd D29(1984)2366.}
\lref\bt{J.H. Brodie and S. Thomas, to appear.}
\lref\zumino{See for example, B.~Zumino,
``Chiral Anomalies And Differential Geometry: Lectures Given At Les Houches,
August 1983,''
UCB-PTH-83/16
{\it Lectures given at Les Houches Summer School on Theoretical Physics, Les Houches, France, Aug 8 - Sep 2, 1983}.}
\lref\kaplan{D.~B.~Kaplan,
``A Method for simulating chiral fermions on the lattice,''
Phys.\ Lett.\  {\bf B288}, 342 (1992)
[hep-lat/9206013].}
\lref\witten{E.~Witten,
``Branes and the dynamics of {QCD},''
Nucl.\ Phys.\  {\bf B507}, 658 (1997)
[hep-th/9706109].}
\lref\skyrme{T.~H.~Skyrme,
``A Unified Field Theory Of Mesons And Baryons,''
Nucl.\ Phys.\  {\bf 31} (1962) 556.}

\lref\am{M.~F.~Atiyah and N.~S.~Manton,
``Skyrmions From Instantons,''
Phys.\ Lett.\  {\bf B222}, 438 (1989).}

\lref\EDSdyons{E.~Witten,
``Dyons Of Charge E Theta / 2 Pi,''
Phys.\ Lett.\  {\bf B86}, 283 (1979).}

\lref\wati{
W.~I.~Taylor,
``Adhering 0-branes to 6-branes and 8-branes,''
Nucl.\ Phys.\  {\bf B508}, 122 (1997)
[hep-th/9705116].}

\lref\igor{
I.~R.~Klebanov,
``D-branes and production of strings,''
in {\it NONE}
Nucl.\ Phys.\ Proc.\ Suppl.\  {\bf 68}, 140 (1998)
[hep-th/9709160].}

\lref\wen{
X.~G.~Wen,
``Nonabelian Statistics In The Fractional Quantum Hall States,''
IASSNS-HEP-90-70.}

\lref\bbst{
B.A.Bernevig, J.~H.~Brodie, L.~Susskind and N.~Toumbas,
``How Bob Laughlin tamed the giant graviton from Taub-NUT space,''
hep-th/0010105.}

\lref\coleman{S.~Coleman,
``Q Balls,''
Nucl.\ Phys.\  {\bf B262}, 263 (1985).}

\lref\llm{
C.~Lee, K.~Lee and H.~Min,
``Selfdual Maxwell Chern-Simons Solitons,''
Phys.\ Lett.\  {\bf B252}, 79 (1990).}

\lref\jlw{
R.~Jackiw, K.~Lee and E.~J.~Weinberg,
``Selfdual Chern-Simons Solitons,''
Phys.\ Rev.\  {\bf D42}, 3488 (1990).}

\lref\khare{}

\lref\bhkk{
O.~Bergman, A.~Hanany, A.~Karch and B.~Kol,
``Branes and supersymmetry breaking in 3D gauge theories,''
JHEP {\bf 9910}, 036 (1999)
[hep-th/9908075].}

\lref\lloy{
B.~Lee, H.~Lee, N.~Ohta and H.~S.~Yang,
``Maxwell Chern-Simons solitons from type IIB string theory,''
Phys.\ Rev.\  {\bf D60}, 106003 (1999)
[hep-th/9904181].}

\lref\koo{
T.~Kitao, K.~Ohta and N.~Ohta,
``Three-dimensional gauge dynamics from brane configurations with  (p,q)-fivebrane,''
Nucl.\ Phys.\  {\bf B539}, 79 (1999)
[hep-th/9808111].}

\lref\ohta{
K.~Ohta,
``Moduli space of vacua of supersymmetric Chern-Simons theories and type  IIB branes,''
JHEP {\bf 9906}, 025 (1999)
[hep-th/9904118].}

\lref\ko{
T.~Kitao and N.~Ohta,
``Spectrum of Maxwell-Chern-Simons theory realized on type IIB brane  configurations,''
Nucl.\ Phys.\  {\bf B578}, 215 (2000)
[hep-th/9908006].}

\lref\hw{
A.~Hanany and E.~Witten,
``Type IIB superstrings, BPS monopoles, and three-dimensional gauge  dynamics,''
Nucl.\ Phys.\  {\bf B492}, 152 (1997)
[hep-th/9611230].}

\lref\ch{
S.~Coleman and B.~Hill,
``No More Corrections To The Topological Mass Term In QED In Three-Dimensions,''
Phys.\ Lett.\  {\bf B159}, 184 (1985).}

\lref\creutz{
M.~Creutz,
``Aspects of chiral symmetry and the lattice,''
hep-lat/0007032.}

\lref\gw{
J.~Goldstone and F.~Wilczek,
``Fractional Quantum Numbers On Solitons,''
Phys.\ Rev.\ Lett.\  {\bf 47}, 986 (1981).}

\lref\cm{
C.~G.~Callan and J.~M.~Maldacena,
``Brane dynamics from the Born-Infeld action,''
Nucl.\ Phys.\  {\bf B513}, 198 (1998)
[hep-th/9708147].}

\lref\mora{
P.~Mora,
``Chern-Simons supersymmetric branes,''
Nucl.\ Phys.\ B {\bf 594}, 229 (2001)
[hep-th/0008180].}

\lref\nishino{
P.~Mora and H.~Nishino,
``Fundamental extended objects for Chern-Simons supergravity,''
Phys.\ Lett.\ B {\bf 482}, 222 (2000)
[hep-th/0002077].}

\lref\kogan{
P.~Castelo Ferreira, I.~I.~Kogan and B.~Tekin,
``Toroidal compactification in string theory from Chern-Simons theory,''
Nucl.\ Phys.\ B {\bf 589}, 167 (2000)
[hep-th/0004078].
}

\def\hf{{1\over 2}}
\def\cs{Chern-Simons}

\def\uu{^}
\def\and{&}
\def\cl{{\cal L}}


\def\LongTitle#1#2#3#4#5{\nopagenumbers\abstractfont
\hsize=\hstitle\rightline{#1}
\hsize=\hstitle\rightline{#2}
\hsize=\hstitle\rightline{#3}
\vskip 0.5in\centerline{\titlefont #4} \centerline{\titlefont #5}
\abstractfont\vskip .3in\pageno=0}

\LongTitle{}{hep-th/0012068}{SLAC-PUB-8729} {D-branes in Massive
IIA} {and Solitons in Chern-Simons Theory}

\centerline{
  John Brodie}
\bigskip
\centerline{SLAC, Stanford University}
\centerline{Stanford, CA 94309}
\vskip 0.3in
\centerline{\bf Abstract}
\bigskip
We investigate D2-branes and D4-branes parallel to
D8-branes. The low energy
world volume theory on the branes is non-supersymmetric
 Chern-Simons
theory. We identify the fundamental strings as the anyons of the
2+1 Chern-Simons theory and the D0-branes as solitons. The
Chern-Simons theory with a boundary is modeled using NS 5-branes
with ending D6-branes. The brane set-up provides for a graphical
description of anomaly inflow. We also model the 4+1 Chern-Simons
theory using branes and conjecture that D4-branes with a boundary
describes a supersymmetric version of Kaplan's theory of chiral
fermions.

\Date{December 2000}

\newsec{Introduction.}

Modeling other physical systems using D-branes is a recent trend
in string theory: black holes, supersymmetric gauge theories,
non-commutative field theory, the fractional quantum Hall
effect, and even the Standard Model
are all examples of systems that have string theory
realizations. It has been known from sometime that
Dp-branes in massive IIA supergravity
have a Chern-Simons term
\eqn\CSgen{\cl_{Dp} = k A \wedge F^{p\over 2}}
on their world volume \ght\ enabling one to model
Chern-Simons theories in string theory. In this paper we will explore
the consequences of having Dp'-branes in the Dp-branes in
massive IIA and take up the task of modeling
anyons, solitons, and edges in the Chern-Simons
theories on the brane world volume theory.

Chern-Simons theories are very interesting for many reasons: one is
that electrically charged particles in the Chern-Simons theory
can have fractional statistics: they are neither bosons nor
fermions, they are anyons.
Anyons have important roles to play in theories of
the fraction quantum Hall effect, HiTC superconductivity,
and more recently in quantum computing.
The Chern-Simons term itself is
related to gauge anomalies in one higher dimension.
It follows from this that at low energies where one can ignore
the kinetic term the 2+1 Chern-Simons theory is
topological.
It is an example of an exactly protected
quantity that does not depend on supersymmetry.
The 4+1 Chern-Simons theory
on a boundary is interesting because it gives
the domain wall fermions studied by Kaplan \kaplan.
Perhaps if we want to solve QCD using string theory,
we should include chiral symmetry in the way
suggested by Kaplan.

In this paper, for a D2-brane in the
presence of a D8-brane, we will identify the 8-2
string ends on the D2-brane as the anyons of the
2+1 non-supersymmetric Chern-Simons theory and D0-branes
outside the D2-branes with $k$ strings attached as
non-topological solitons with electric charge k. By
introducing NS5-branes we can add a boundary to our D2-brane and
study the chiral theory on the boundary. We identify the massless
chiral fermions as the light strings on the 2+1 boundary. This
provides for a microscopic picture of charge inflow: long strings
move to the boundary creating massless charged particles from
``nowhere'' and therefore violating charge conservation in the 1+1
dimensional theory. For a D4-brane the situation of a D8-D4-D0 is
supersymmetric. The D0-brane is an instanton carrying $k$ units of
electric charge in the D4-brane
world volume. Putting a boundary on the D4-brane induces chiral fermions
in 3+1 and reminds one of Kaplan's theory of chiral symmetry on
the lattice \kaplan\creutz.

The outline of the paper is as follows:
In section 2, we will review the string creation process
which happens when a D0-brane crosses a D8-brane and
go over the argument of \bdg\ as to why this
is related to an anomaly inflow of a  1+1 dimensions.
In section 3, we will relate the anomaly inflow of a 3+1
gauge theory to a string creation when a D2-brane crosses a D8-brane
carrying magnetic flux. The magnetic flux is
produced by a D0-brane inside the D2-brane. We identify
the $k$ strings created as $k$ anyons of the 2+1 Chern-Simons
theory and the D0-brane
as a soliton carrying $-k$ units of electric charge.
In section 3.8 we propose that a $U(M)$ level $k$
non-Abelian Chern-Simons
theory is dual to the fractional quantum Hall effect
at level $\nu = {k\over M}$. This proposal is based a
T-duality relation between the brane scenario discussed in
this section and the brane scenario of \bbst.
In section 3.9, we show how to put the Chern-Simons
theory on a boundary by introducing NS5-brane on
the edge of the D2-brane. We then show how
anomaly inflow from the 2+1 theory into the
chiral theory on the 1+1 boundary is realized in
this string model. In section 4, we generalize the
results of the 2+1 dimensional Chern-Simons theory to 4+1 dimensions
and propose that they are related to the
domain wall fermion proposal used in lattice QCD \kaplan\creutz.
In section 5, we will discuss some meta-stable non-supersymmetric
bound states of D2-branes and D8-branes.

There are several papers on supersymmetric Chern-Simons
theories in 2+1 realized on branes of type IIB \koo\ohta\ko\bhkk\ as well
as papers on solitons \lloy.
The novelty here is that the Chern-Simons coefficient has
as different origin; it comes from the D8-branes. Moreover,
the Chern-Simons theory on the boundary in easily studied in
this brane construction, it is related to
string creation of \hw, and this construction
has a close relationship with the fractional quantum Hall
construction of \bbst\ as will be discussed.

For other occurences of Chern-Simons theory in 
string theory see \kogan\nishino\mora.

\newsec{String creation by a D0-brane crossing a D8-brane:
A review of the supersymmetric case.}

\subsec{A few comments about Massive IIA supergravity.}
Massive IIA has a positive cosmological constant
proportional to $(F^{(0)})^2$, the dual
of the ten form field strength that couples
to the D8-brane, and a linear
dilaton potential.
There is no known ulta-violet definition of massive IIA supergravity.
The problem is that the
string coupling constant grows stronger the farther one is away from
the D8-brane
\eqn\LinDil{{1\over g_s} = {1\over g_0} - |r-r_0|^{5/4}}
At $r=r_0$, $g_s = g_0$ and then grows until $r=r_0 + ({1\over g_0})^{4/5}$
where the coupling blows up. However we can make this distance
(measured in string units) very big by taking $g_0$ to be very small.
Then we can work in the
supergravity regime near the
D8-brane and not venture out to where gravitational
effects become strong.

If one doesn't feel comfortable with the Massive IIA supergravity
solution one could introduce orientifold planes
which will cut off the growth of the dilaton
before one looses control. One should just make sure
to arrange the probe brane (i.e. D0,D2,D4-brane) such that
there are unequal numbers of D8-branes one the left and right hand
side of the probe. This will insure that the Chern-Simons
coefficient does not vanish since as we shall see
\eqn\coef{ {\tilde k\over 2} = {k -16\over 2}}
where $k$ is the number of D8-branes and
where the $-16$ comes from the orientifold plane charge.
In such a theory the coefficient of the
Chern-Simons theory on the probe brane $\tilde k$
is always integer.

Massive IIA supergravity is called ``massive'' because
the antisymmetric tensor field $B_{\mu\nu}$ gets a mass by eating the
RR-vector field $C_{\mu}$. One can understand this by writing a
new gauge invariant RR field strength $\tilde G$ as
\eqn\Gfield{ \tilde G  = G + kB}
where $G = dC$. Expanding out the kinetic term
for $C$ in the IIA action, we find a mass term for $B$
\eqn\expand{\cl_{IIA} = \tilde G * \tilde G = G*G + k B*G + k^2 B*B}
as well as a coupling of $B$ to $dC$.

\subsec{Anomaly in 1+1 dimensions.}
\subseclab{\oneplusone}

Let us consider two D5-branes on type IIB string theory
extending in directions 012345 and 016789.
As described in \bdg, the low energy theory on the
two dimensional intersection is a $U(1)\times U(1)$ field
theory with $N=(0,8)$ supersymmetry. There is a chiral
spinor charged as $(1,-1)$
under the gauge group which is the 5-5 string.
Notice that
one cannot give a mass to the chiral
fermion since in the brane set-up that involves
separating the D5-branes and stretching the 5-5 string
and there is no such direction
where this is possible. This theory has an anomaly at
one loop if one turns on
a background gauge field strength $F_{01}$
since the current obeys the relation
\eqn\index{\dd^{\sigma} J_{\sigma}^3 = F_{\mu\nu}\epsilon^{\mu\nu}}

and is therefore not conserved when
$F_{01} \neq 0$.
Because the 1+1 dimensional gauge theory is anomalous
but the ten dimensional bulk theory is anomaly free,
there must be inflow of charge that cancels the anomaly
and restores gauge invariance.
This inflow comes from the D5-branes.
As was noted in \ch\ the inflow current is
perpendicular to the electric field.
This is reminiscent of the quantum hall effect
which we shall make more precise later
in section 3.8.

\subsec{Chern-Simons theory in 0+1 dimensions.}

\ifig\tad{Integrating out the fermions in 0+1 induces a
Chern-Simons term.}
{\epsfxsize2.0in\epsfbox{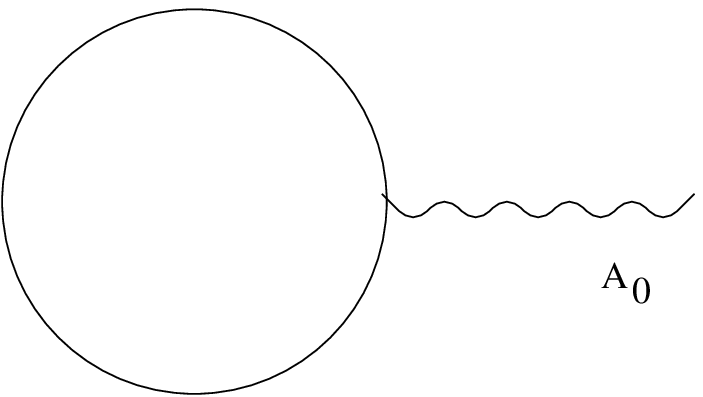}}

Upon T-duality along 12345, the theory discussed in section
\oneplusone\ becomes a D0 brane and a D8-brane.
There is a $U(1)$ gauge
symmetry on the D0-brane and a $U(1)$ flavor symmetry
on the D8-brane since the 8-brane coupling is
very weak in the infra-red. Virtual 0-8 strings
are fermions with a real mass $<\phi > = m$,
the separation distance between the 0-brane and the 8-brane.
Doing a 1-loop calculation and integrating out the
massive fermions induces a Cherm-Simons
term and a potential for the scalar $\phi$.
\eqn\csZero{\cl_{D0} = -\hf {m \over |m|} (A_0+ \phi)}
The 1-loop term in the
open-string channel is dual to a tree-level
term in the closed string channel.
Due to a non-renormalization theorem
potential is the same
whether calculated in
supergravity or in field theory \dkps\bss.
In fact we can do the supergravity calculation as follows:
The metric for a Dp'-brane is given by
\eqn\metric{ds^2 = f(r)^{-1/2}dx^2 + f(r)^{1/2}dy^2.}
where $x^\mu$ $\mu=0...p'$ are the coordinates parallel to the
brane, $y^a$ $a=p'+1...9$
are the coordinates perpendicular to the
brane, and $r = \sqrt{y^ay^a}$.
The dilaton obeys  the equation
\eqn\dilaton{e^{-2\phi} = f^{3-p'\over 2}.}
A parallel probe Dp-brane action is
\eqn\braneAction{S = e^{-\phi} \int \sqrt {\det G}}
where $G_{\mu\nu}$ is the metric on the Dp-brane
induced by the metric in the bulk \metric.
Plugging \metric\ and \dilaton\ into \braneAction\ we find
that the potential for the scalar field on the
brane is
\eqn\potential{V(\phi) = f^{p'-p-4\over 4}(\phi)}
where the scalar field is related to the coordinates
via the relation $\phi = M_s^2 x$.
For a D8-brane and a D0-brane we find that
\eqn\potDeightZero{V(\phi) = f(\phi) = -\phi}
Notice that there is a difference in a factor of
$\hf $ between equation \potDeightZero\ and \csZero.
This is because of a different normaliztion conversion:
In supergravity it is assumed that the value of
$F^{(0)}$ jumps from 0 to 1, but in
field theory it is assumed that there is a jump from $-\hf$ to
$\hf$. This is related to the
notion of a ``half-string creation'' - see \igor\ for discussion.

\ifig\zeroeight{When a D0-brane crosses $k$ D8-branes, $k$  fundamental
strings are created.}
{\epsfxsize2.0in\epsfbox{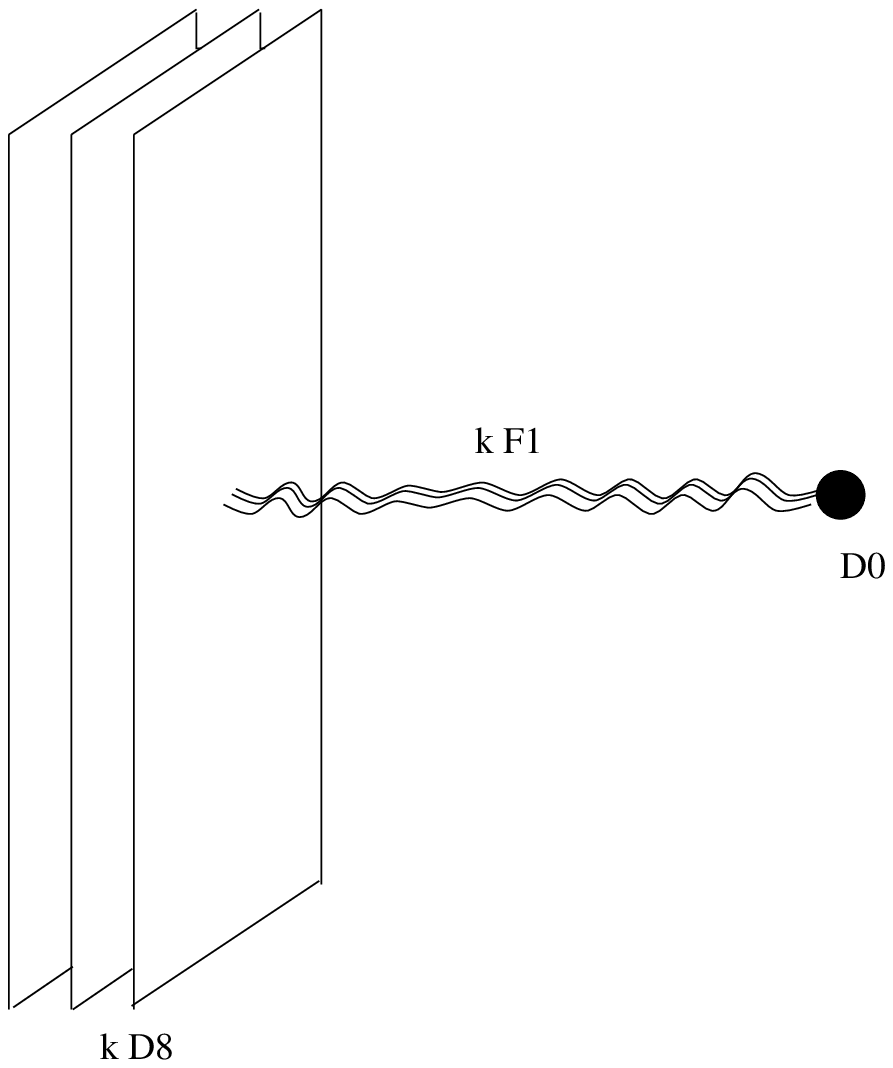}}

As the D0-brane crosses the
D8-brane the chirality of the
fundamental field changes sign.
Therefore \csZero\ also changes sign.
To prevent such a discontinuous jump,
we must introduce some charge that
can compensate for this transition.
This is supplied by the fundamental
string that is created.
In fact, the electric flux that one must
turn on in the 1+1 theory to induce
an anomaly becomes momentum of the
D0-brane transverse to the D8-branes\bdg.

\subsec{Charge conservation in ten dimensions.}
\subseclab{\SecCharge}

Another way one can view the brane creation is through
supergravity \strominger\az. Consider the action for the antisymmetric
field which couples to the string.
\eqn\Bfield{\cl_{IIA} = H\wedge *H + B^{(2)}\delta^{(8)} +
k B^{(2)}G_{RR}^{(8)}.}

Where the delta function is the string source and the 8-form field
strength couples magnetically to the D0-brane. The coupling of the
antisymmetric tensor field to the 8-form field strength is
peculiar to massive IIA supergravity \romans. The equations of
motion following from this action are \eqn\eom{d*H + \delta^{(8)}
+ k G_{RR}^{(8)} = 0.} Because flux lines have no where to go on a
sphere \eqn\sphere{\int_{S^8} d*H = 0 = \int_{S^8} (\delta^{(8)} +
k G_{RR}^{(8)}) = (Q_{NS} + k Q_{DO})} If we didn't have the
coupling in \Bfield\ charge conservation would have been violated.
This is the reason that a single string can end on a D0-brane only
in massive IIA supergravity and not in ordinary Type IIA
\strominger.

\newsec{String creation by a D2-brane crossing a D8-brane:
The non-supersymmetric case.}

\subsec{Anomaly in 3+1 dimensions.}
\subseclab{\anomalyFour }

Now lets consider two D6-branes in type IIA string theory
along directions 0123456 and 0123789. The theory on the
intersection is a 3+1 dimensional $U(1)\times  U(1)$
gauge theory with a single chiral fermion with charge
$(1,-1)$ under the gauge group. In terms of branes the
chiral fermion comes from the 6-6 strings. Again one cannot
give the chiral fermion a mass in the brane picture
since one cannot separate the 6-branes. This theory is
non-supersymmetric, and one can think of it as coming
from a supersymmetric theory with a FI D-term \kMc\bdl.
There is a gauge anomaly from the one-loop
triangle diagram which is cancelled by inflow from
the 6+1 dimensional theory off the intersection.
However, according to
\eqn\indexThree{\dd^{\sigma} J_{\sigma}^5 = \epsilon^{\mu\nu\rho\lambda}F_{\mu\nu}F_{\rho\lambda}}
there will be no anomaly unless there
is a background field where $F\wedge F$ is non-zero.
To satisfy this we must have non-zero electric field
$F_{01}$ and non-zero magnetic field $F_{23}$.
In terms of branes the electric flux is a fundamental
string in directions 03 bound to the D6-brane
while the magentic flux is a
D4-brane in directions 03456
parallel to and bound to the D6-brane in directions
0123456.

Another way to induce an anomaly in 3+1 dimensions
is to put an instanton in
the 3+1 theory. Such an instanton
could be provided by placing a
Euclidean D2-brane inside the
D6-brane along spatial directions 456.

\subsec{Chern-Simons theory in 2+1 dimensions.}
\subseclab{\csThreeSec}

\ifig\wave{Integrating out fermions induces a Chern-Simons term in
2+1 dimensions.}
{\epsfxsize2.0in\epsfbox{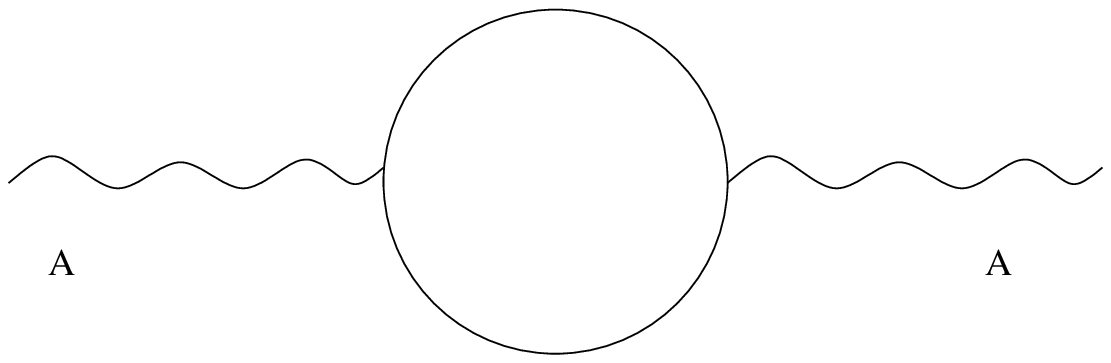}}

Now lets T-dualize the brane configuration along directions 3456
and in doing so dimensionally reduce the theory on the
intersection to 2+1. \foot{One could equivalently consider two
D5-branes intersecting over a 2+1 surface in IIB.} The system is
now a D2-brane along 012 and a D8-brane along 012456789. In
addition to the usual fields of $N=8$ super Yang-Mills in 2+1
dimensions, the gauge field, seven scalars, and eight fermions all
in the adjoint representation, there is a parity violating
fundamental fermion that couples to the real adjoint scalar
corresponding to the 3-direction. \eqn\yukawa{ \cl_{D2} = X
\bar\psi \psi + ...} Setting $<X> = m$, at one loop, after
integrating out the massive fermion there is a \cs\ term induced.
\eqn\csThree{\cl_{D2} = \hf {m\over |m|} \epsilon^{\mu\nu\rho}
A_{\mu} F_{\nu\rho}.} This one-loop term is not renormalized \ch\
and therefore we conjecture it to be the same as the term
calculated using supergravity. $k$ D8-branes give a Chern-Simons
coefficient ${k\over 2}$. The non-renormaliztion of the
coefficient of \csThree\ is related to D8-brane charge
conservation in the bulk theory. Interestingly, as was mentioned
in section 2.1 in a IIA theory with O8-planes as well as D8-branes
the Chern-Simons coefficient is always integer. This agrees with
the fact that there is an anomaly in the compact $U(1)$ field
theory with even Chern-Simons coefficient \redlich.

There is a different 1-loop diagram that contributes to
the Coleman-Weinberg potential. For $X << M_s$
it is
\eqn\cwPot{V(X)_{FT} = -|X |^3 \log(X^2 )}
In the other limit, $X >> M_s$ one
cannot use field theory because all of the
stringy modes become relevant, but one can, by channel duality,
 use supergravity. Using the formulas
\metric\ \dilaton\ and \braneAction\ in the
previous section, one finds that
the potential for the 2+1 dimensional theory on the
D2 brane is
\eqn\sugraPot{V(X )_{SUGRA} = kf^{\hf } = -k X^{\hf } }
where $k$ is the number of D8-branes.
The low energy theory on the $N_c$ D2-branes is then
a $SU(N_c)$ gauge theory with 8 gauginos and 6 scalars
all transforming in the adjoint representation.

\subsec{D0-branes in D2-branes with D8-branes.}

Now let's consider putting a D0-brane inside the D2-brane
in massive IIA
\foot{In much of this discussion we will imagine holding the
D2-brane fixed. This is accomplished with NS5-branes and will be
discussed in section 3.8}. We will not attach strings to the D0-brane
and we will see in the next section that there is no
violation of charge conservation.
The potential $C^{(1)}_{RR}$ that couples
to the D0-brane couples also to
the gauge field on the brane $F$ through
the coupling
\eqn\DBI{\cl_{D2} = C^{(1)} \wedge F^{(2)}.}
Therefore, the D0-brane charge appears in the 2+1 world
volume as magnetic flux.
Moreover, because of the Chern-Simons
term in the Maxwell Lagrangian on the D2-brane \csThree\ we see
that the magnetic flux induces $k$ units \foot{Here
we must be careful to note that
we have renormalized $k$ such that the Chern-Simons coefficient
is twice the eightbrane charge.} of
electric charge
\eqn\EOMcsThree{d*F = k F.}

From string theory, typically one expects that the D0-brane
dissolves inside the D2-brane.
This is because in IIA string theory there
is an instability due to a  tachyonic mode on the 0-2 string.
However, we will argue that
this tachyon does  not have to roll off to infinity,
but can be  in fact stabilized by the 1-loop
term coming from integrating out the 0-8 strings which
induce the linear scalar potential, the superpartner
of the Chern-Simons term \csZero.
The potential for the D0-brane is
\eqn\ZeroPot{V_{D0} = -k\phi + \phi^2q^2 - \mu^2q^2+....}
where $m$ is the tachyonic mass of the
0-2 string.
The equations of motion following from \ZeroPot\ are
\eqn\eomPot{\eqalign{-k + \phi q^2 \and = 0 \cr
(\phi^2 - \mu^2)q \and = 0}.}
Equations \eomPot\ have two minima: one at $q = 0$ with $\phi = \infty$
the other at
$q = \infty$ with $\phi = 0$.
The solution with $q = \infty$ corresponds to the
D0-brane spreading out while $q =0 $ corresponds to a
confined D0-brane.
Physically the latter solution corresponds to
the D0-brane being pushed away  from the
D8-brane by the linear potential.
However,
the D0-brane cannot leave the D2-brane or it will violate
charge conservation.
The D0-brane is stuck to the D2-brane
but repelled from the D8-brane. This causes the
D2-brane to bend. The scalar field on the
D2-brane satisfies Laplace's equation
\eqn\laplace{\nabla^2 X = \delta^{(2)}(x).}
This equation has a logarythmic solution
in 2+1 dimensions. These vortices are therefore
charged under the electric field, the Higgs field,
and carry one unity of magnetic flux.
There are three forces now on the D0-brane: repulsion due to the D8-brane,
attraction due to the bending of the D2-brane, plus an attraction due to the D2-brane. When the D0-branes is far from the D2-brane,
the attractive bending cancels the D8-brane repulsion,
and so the D2-brane attraction wins. When the D0-brane is inside the
D2-brane, the D8-brane repulsion dominates since there is no bending
and the D0-brane cannot feel the D2-brane force.
We claim then that the forces balance when the D0-brane is
a string length from the D2-brane.
\foot{It is interesting
to note that the mechanism of relocalizing the D0-brane inside the
D2-brane can be understood as a 1-loop effect
in 2+1 field theory: There first are massive fermions and
a magnetic flux, then we integrate out the fermions which
generates a Chern-Simons term giving the gauge field a mass,
the magnetic flux then confines into a magnetic vortex.
This scenario is similar to the confinement by instantons mechanism
proposed in \brodie. There Euclidean D0-branes were argued to
confine strings inside D4-branes.}

\subsec{Non-topological solitons}

Its clear that the D0-branes are non-topological
solitons of the Chern-Simons theory. Topological vortices are
typically associated with a broken $U(1)$ gauge invariance,
but here the gauge symmetry is preserved. If fact there are no massless
fundamental scalar fields with which we could Higgs the gauge field.
There are however in Chern-Simons theories stable non-topological
solitons which are like the Q-balls of \coleman.
In most gauge theories flux wants to spread out since spreading
lowers
the energy. One can see  this from the formula for the energy
\eqn\energy{{\cal E} = V(E\uu2 + B\uu2)}
where $V$ is the volume of the soliton
and $E$ is the electric field and $B$ is the magnetic field. If we take the flux to
be constant
\eqn\flux{\Phi = BV = N}
then the equation for the energy becomes
\eqn\energy{{\cal E} = {N\uu 2 \over V}} Therefore the more spread out
the object, the lower its energy.
This is what one is familiar with
in string theory: the D0-brane wants to spread out in the
D2-brane.
This will not be the case if the soliton is also charged under some
scalar field since the energy of the scalar field will grow like the
volume of the soliton.
Therefore once the D0-brane is charged under the neutral
scalar field $X$,
its size will be stabilized.
The neutral scalar field corresponds to motion of the D2-brane
transverse to the D8-brane which we can make massive
by introducing more branes (as will be explained later in
section 3.9).
Presumably $X$ will have a solution such as
\eqn\solnPhi{X(r) = \log r}
since it satisfies \laplace.
Clearly this solution does not have finite energy since $X$ does
not approach zero at $r = \infty$ where $r$ is the spatial coordinate
on the D2-brane. However, the total system does have finite
energy because there are
strings in the theory charged oppositely under the scalar field and
so the total scalar charge is zero just like the total electric
charge is zero. The logarythmic solution
\solnPhi\ will turn into multipole solution
which has a power law form and dies away at infinity.
In this way, the non-topological solitons that
we find in this string theory set-up appear to be different from the
non-topological solitons discussed in the literature
since there a single soliton has finite energy \llm\jlw.

\subsec{Topological solitons}

Since we have discovered that we can have
non-topological solitons in our brane model of
the Chern-Simons theory, it is natural to ask
whether we can have topological solitons as well.
To have topological solitons we must introduce some
extra branes that will play the role of the fundamental
scalar field $q$. We will choose to add D4-branes.
\foot{We could have also used D6-branes but that adds some complications
in massive IIA as we will see later.}
The boundary conditions on the 2-4 strings
allow for one tachyonic scalar field
when the D2-branes are on top of the D4-branes, and
what's more
we already know that the neutral adjoint scalar field
has a tadpole from \sugraPot. The potential for the D2-brane
can then be modeled as we did for the
D0-brane \ZeroPot\ by the following equation
\eqn\twoBranePot{V_{D2} = -\mu^2q^2 - X^{\hf} + X^2 q\uu2}
This solution again has the following minima:
If $q$ gets a vacuum expectation value and  $X = 0$, the
the $U(1)$ gets broken and there can be topological solitons.
In the brane
picture the D2-brane will dissolve into the D4-brane
while the D0-branes remain of finite size since they are
instantons in the D4-branes.
On the other hand another minimum is when $q = 0$
and $X$ gets a vacuum expectation value.
This will lead to the non-topological solitons that we discussed above.
In the brane picture, the D8-brane pushing on the D2-D4 bound state
will cause the D4-brane to bend
while the D8-brane pushing on the D2-D0 bound state
 will cause the D2-brane to bend.
Both the D2-brane and the D0-brane will be undissolved.

\subsec{Charge conservation in ten dimensions with a
D2-brane.}
\subseclab{\SecChargeTwo}

As in section \SecCharge\ we can look at the equations
of motion following from the massive IIA action
and see how charge conservation is satisfied.
The action is the same as \Bfield\ except for
a new coupling of the B-field to
the gauge field on the brane $F$.
\eqn\BfieldTwo{\cl_{IIA} = H\wedge *H + B^{(2)}\delta^{(8)} +
k B^{(2)}G_{RR}^{(8)} + B^{(2)}\wedge *F^{(1)}\delta^{(7)}.}
The equations of motion are
\eqn\eomTwo{d*H + \delta^{(8)} + k G_{RR}^{(8)} + *F^{(1)}\delta^{(7)} = 0.}
Now when we integrate over the 8-sphere we find
\eqn\sphereTwo{Q_{NS} + k Q_{D0} + \int_{S^1} *F = 0.}
Notice that this equation can be satisfied in
many different ways:
1) The D0-brane charge can cancel the
electric charge on the brane. This solution
is a single vortex in 2+1 dimensions.
\eqn\vortexCharge{k Q_{D0} + \int_{S^1}*F = 0}
2) The string charge can cancel the
electric charge on the brane.
This solution is an electron in 2+1 dimensions.
\eqn\electricCharge{Q_{NS} +   \int_{S^1} *F = 0}
3) The D0-brane charge cancels with the
D2-brane charge. There is no electric charge in the
2+1 world volume.
\eqn\zeroCharge{Q_{NS} + k Q_{D0} = 0}
We conclude that a D0-brane in a D2-brane in massive
IIA behaves like a cut string. The cut string is
sensible solution if it has its end on a D2-brane, but
not is if has its end in the bulk:
Likewise, the D0-brane does not violate charge conservation
if it is inside a D2-brane in massive IIA, but it cannot
go outside of the D2-brane.

\subsec{Non-supersymmetric String creation}

Because the real mass of the fundamental field
changes sign from one side of the D8-brane
to the other, the sign of \csThree\ also
changes sign. However, if there is no
background field strength $F_{\mu\nu}$, then
\csThree\ is zero, and there is no problem.
However, if $F_{\mu\nu}$ is non-zero,
then there is a problem with charge conservation
because equation \EOMcsThree\ say that the D0-brane
carried $k$ units of
electric charge when on one side of the D8-brane
but $-k$ units of electric charge on the other side.
The difference in electric charge must be compensated for
by a string creation.
One way to have a non-zero field strength $F_{\mu\nu}$
on the brane is, as we saw in section 3.3 , to bind a D0-brane
to the D2-brane. The D0-brane is
exactly T-dual to the D4-brane from the
previous section 3.1
which played the role of magnetic
flux inside the D6-brane.
When the D2-D0 system crosses the
D8-brane, a string is created.
Notice that the momentum of the
D2-D0 bound state in the 3-direction
is T-dual to
the electric field
that was necessary to have an anomaly in
3+1 dimensions.

In the previous section we said that we could induce charge inflow
into the anomalous 3+1 theory by adding an instanton. Now we will
show that upon dimensional reduction to 2+1 dimensions the
instanton implies charge inflow. T-dualizing the D6-D6
configuration with a Euclidean D2-brane along directions 3456 we
find a D2-D8 configuration with a Euclidean D0-brane along the
3-direction. How do we think about a Euclidean D0-brane? The
analogy comes from open-closed string duality in string theory.
There events such as the creation and annihilation of charged
particles, i.e. loops are represented in the closed string channel
as a propagating particle. However we are use to thinking of a
propagator as a particle with some momentum moving between two
sources. So let think of the Euclidean D0-brane in the same way.
The Euclidean D0-brane is a D0-brane that moves along the
3-direction. Since there is an attractive force between the
D2-brane and the D0-brane, they will form a bound state and move
together. Therefore we have a D2-brane with a D0-brane bound to
it, moving in the 3-direction just as described above. When the
D2-brane crosses the D8-brane a string is created. This is the
T-dual of the charge inflow in 3+1 dimensional anomalous theories.

\subsec{Strings as anyons.}

In the previous section \csThreeSec\ we saw that integrating out
fermions in 2+1 QED leads to a Chern-Simons term.
Because of the equations of motion \EOMcsThree, whenever we have one unit of
magnetic flux we have
$k$ units of electric charge.
Conversely when we have
1 unit of electric charge we can have $\ovk$ units
of magnetic flux. That is fractional
magnetic flux. Moreover when one circles such
fractional magnetic flux with an electrically
charged particle on picks up
a fractional phase due to the Bohm-Ahramov effect \CSreview.
\eqn\phase{e^{2\pi ie\int A\cdot dx} = e^{2\pi i{e^2\over k}}}
These are the anyons of condensed matter physics with
spin $\ovk$. In the brane set-up the ends of the
strings are electrically charged. According to the Chern-Simmons equations
of motion
this implies that each string end should carry
$\ovk$ units of magnetic flux.
If the D0-brane
crosses  $k$ D8-branes it will have  $k$ strings
ending on it. Therefore, a D0-brane
and $k$ strings have exactly opposite charge and flux.
The strings are the anyons and the
D0-branes have the quantum numbers of  a composite of $k$ anti-anyons.
In fact, if one begins with the D0-brane to the D8-brane-side of the
D2-brane and pulls the D0-brane through the D2-brane, one
will have $k$ strings on either side of the D2-brane.
If one now moves the D0-brane down, the strings on the
D8-brane side can all move independently while the
strings on the D0-brane side are all bound to
the D0-brane. In this way, the soliton is
literally a composite of $k$ anyons.

\subsec{Chern-Simons as dual to the fractional quantum Hall
effect.}

In \bbst\ a brane model for the fractional quantum Hall effect
involving $M$ D0-branes, $N$ D2-branes and $k$ D6-branes
was presented. The system of $N$ D0-branes, $M$ D2-branes, and $k$ D8-branes
presented in this paper
is T-dual along the spatial directions of the
D2-branes to the configuration of \bbst.
In terms of the low energy effective field theory this
duality implies that
an fractional quantum Hall effect with
filling factor $\nu = {k\over M}$ is dual to a $U(M)$ Non-Abelian
theory
with Chern-Simons level $k$.
One should think of the D2-D0-D6-brane theory on a $T^2$
rather than the $S^2$ of \bbst. If, for example,
the filling factor were ${1\over 3}$,
then there is one D6-brane, one string, and 3 D0-branes
per cell defined by the periodic torus.
This torus here is playing the role of
the periodic lattice in condensed matter systems.
This is then T-dual to a $U(3)$ theory
on a similar periodic lattice with the D0-branes
playing the role of the compensating negative charge.
The atomic nuclei in condensed matter systems.

It is amusing that this relates the FQHE to
a non-Abelian Chern-Simons theory where the electrons
carry a non-Abelian index. This suggests that in this
picture the electron has structure much as
in composite fermion proposals of \wen.

\ifig\tad{Long 8-2 strings want to move to the boundary, turn into
6-2 strings and become light. This provides the current which
flows into the 1+1 theory on the boundary of the Chern-Simons
theory making it anomalous. A D0-brane outside the D2-brane has
$k$ strings ending on it. The D0-brane behaves like a soliton with
electric charge $k$ and magnetic flux 1. The string ends on the
D2-brane behave like anyons of electric charge 1 and magnetic flux
$\ovk$.} {\epsfxsize3.0in\epsfbox{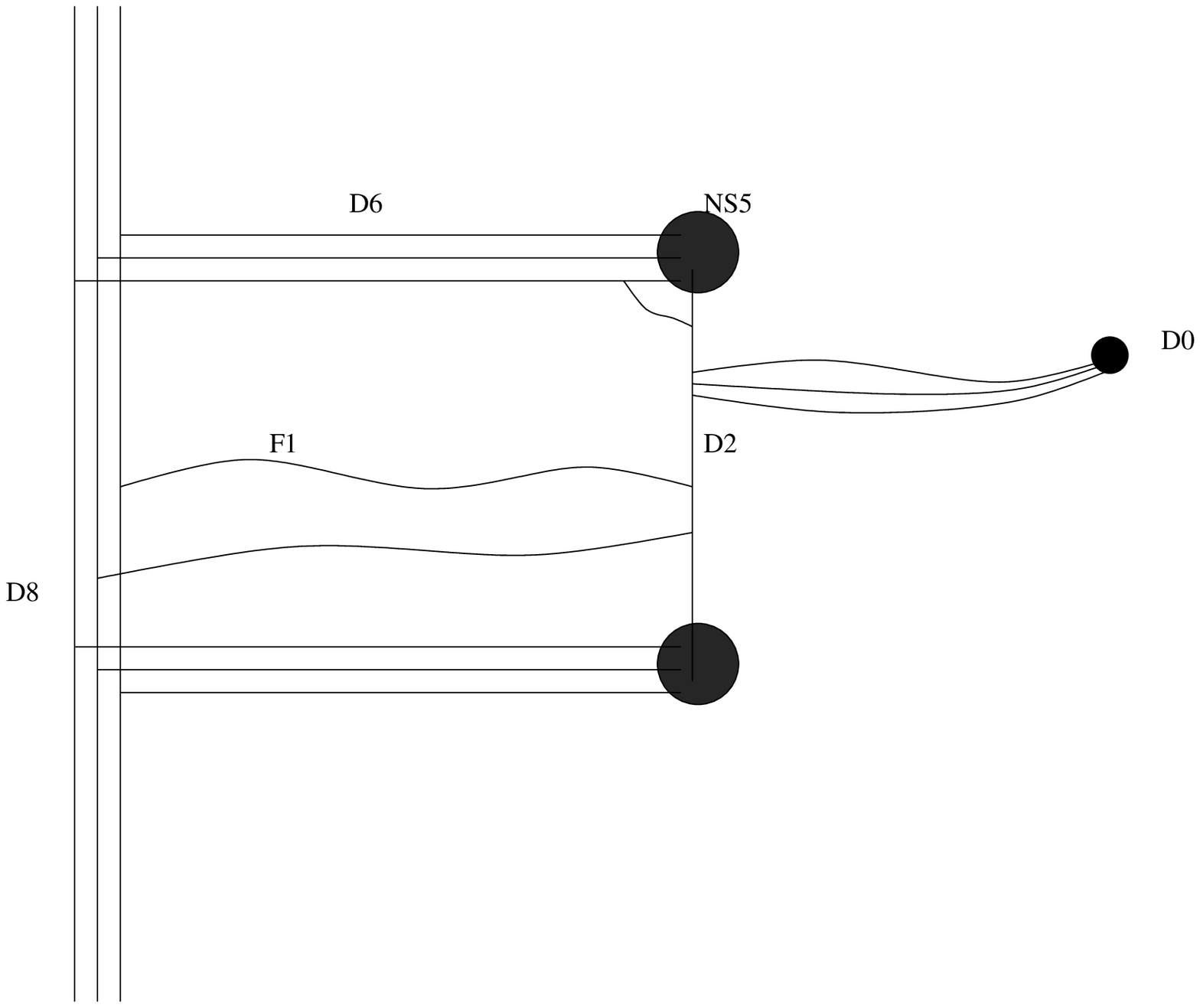}}

\subsec{Chern-Simons theory on a boundary.}
One can make a brane set-up to model the Chern-Simons
theory on a boundary.
We simply have the
D2-brane end on NS 5-branes. Consider D2-brane in 012, D8-branes
in 012456789 as before. Now add NS5-branes along 014567.
The NS5-brane create a boundary on which the D2-brane
can end. However, as with a D0-brane in massive IIA
the coupling in equation \Bfield\ demands that
a D6-brane end on the NS5-branes to conserve
NS charge. Therefore, along the boundary of the D2-brane
there will be
D6-branes in the 0134567 directions.
The massless 2-6 strings are in fact
chiral fermions because the boundary conditions for
a 2-6 strings is still 6ND, the same boundary conditions
as for the 2-8 strings; the strings can move
from the 8-brane onto the 6-brane without changing
their boundary conditions.
We know that the 2-8 strings are parity violating
in the 2+1 theory since they come from
a dimensional reduction
of the chiral 3+1 theory discussed in
section 3.1.

The chiral theory on the right edge of the
D2-brane carries $k$  right moving massless chiral fermions
while on the left edge there are  $k$ left-moving chiral
fermions. If the boundary were a circle, then
the right movers would go around and turn into left movers.
Note that the left and
right moving 2-6 strings {\it cannot} join to make a 6-6 string
that is neutral under the $U(1)$ gauge field
on the D2-brane.

The 2+1 Chern-Simons theory is not gauge invariant on
the boundary. One can see this by starting from the
Lagrangian
\eqn\csLag{\cl_{D2}  = k A_2F_{01} + J^2A_2 + ...}
If we now take consider a small variation
in $A_2$
\eqn\puregauge{\delta A_2 = \dd_2 \alpha}
Under this variation we find a surface term
\eqn\surface{\alpha(k F_{01} + J^2)}
For this surface term to vanish means that
\eqn\anomaly{k E_1 = J^2}
which states that electric field along the boundary
is proportional to current moving transverse to the
field in the bulk. This is very much like the Hall effect.
Moreover, equation \anomaly\ is just the anomaly equation
in 1+1 \index\ if we identify
\eqn\newcurrent{J^2 = \dd_0 K^{02}}
where $K^{\mu 2}$ is the ``axial'' current in
1+1 theory.
This suggests that there should be a
1+1 gauge theory with $k$ chiral fermions on the boundary.

How is the 1+1 anomaly equation \anomaly\ manifest in the
brane construction? The massive 8-2 strings in the bulk
move to the boundary where they become massless 6-2 strings
which then begin to move along the boundary at the speed of light.
From the 1+1 point of view it appears that charge was created
from nowhere and that there is an anomaly. However, from the 2+1 point of
view charge entered the boundary from the bulk and there
is no violation of charge and no anomaly. This is a microscopic
picture in terms of strings of the inflow mechanism of \ch.

Another attractive feature of having
a boundary for the D2-brane
is that this lifts the scalar field
with the runaway potential \sugraPot.
Therefore, the 2+1 field theory is stabilized, at
least perturbatively. Likewise, the supergravity
theory is stabilized for $g_s << 1$.
This is very similar to the set-up discussed in
\bt.

\subsec{Chiral anomaly on the boundary dimensionally reduced to 0+1.}

We can consider dimensionally reducing the above
brane configuration describing a 2+1 Chern-Simons theory with
a boundary. This amounts to
T-dualizing the branes along the 1-direction.
The D2-brane becomes a D1-brane extending in the 2-direction
suspended between the two NS5-branes.
The anomalous boundary theory is then 0+1 dimensional
Chern-Simons.
The D6-brane ending on the NS5-brane turns into a D5-brane
extending along directions 034567 which
ends on the NS5-brane in directions 014567.
The 5-branes then join into a
three 5-brane junction involving an
D5, NS5, and $(1,1)$ 5-brane extending in 45 degrees along
the 13-direction.
The D1-brane can now move along the NS5-brane.
When the D1-brane encounters the three 5-brane
junction it can now move onto a different branch.
As it crosses onto the $(1,1)$ 5-brane it
will must turn into a $(1,1)$ 1-brane.
This means a fundamental string must be
appear.
The inflow of the current flowing into the 1+1 boundary theory
maps to
the D1-string turning into a $(1,1)$-string
under T-duality.

\newsec{D0-D4-D8. Five dimensional Chern-Simons theory.}

Consider a system of D4-branes along directions 01245, D8-branes
along direction 012456789, and D0-branes marginally bound to the D4-brane.
This theory is similar to the case with a D0-D2 bound state
in the background of a D8 brane, except that it is supersymmetric.
In this case which was analyzed in \bdg, the D0-brane is an
instanton in a 4+1 dimensional worldvolume theory. Due to the
D8-brane there is a 5d Chern-Simons coupling
\eqn\fiveCS{{\cal L}_{D4} = F\wedge *F + k A\wedge F\wedge F.}
Since both terns in the action \fiveCS\ have two derivatives
we cannot ignore the kinetic term as we did in the 2+1
Chern-Simons theory.

If $F\wedge F$  is non-zero and the D4-brane crosses
the D8-branes, a string must be created to maintain
charge conservation on the D4-brane.
 Because of the Chern-Simon term \fiveCS\ the
instanton carries electric charge. What happens when the
instanton gets fat?
We'll conjecture that the instantons cannot get fat in
4+1 Chern-Simons theory. From the 0+1 gauge theory point
of view this means that the Higgs branch is lifted.
The way that this happens is similar to the way
magnetic flux became confined in the previous
section.  Separating the 8-0 strings from the D0-brane
on putting them far away on the D4-brane
for the moment, from the point of view
of the $0+1$ dimensional theory living on the
D0-brane there is the following potential:
\eqn\QMpot{V_{D0} = k \phi + \phi^2q^2 + ...}
where we have only written the terms involving
$\phi$.
The minimum solution to
\QMpot\ is $q = 0$ and $\phi = -\infty$.
Because of the
linear potential for $\phi$, the
Higgs branch of the 0+1 dimensional
theory is lifted.
This says that, the D0- branes want to shrink
to zero size inside the D4-brane and leave but
it cannot since it carries
electric charge. So the D4-brane bends in response to
the force in the D0-brane due to the D8-brane. One can picture this as the
D0-brane pulling off little one dimensional pieces of the D4-brane.
However, as was shown in \cm\ these pieces are just
fundamental strings. The final picture is then the D0-brane outside
the D4-brane with $k$ strings attached. From the instanton
moduli space point of view, the
D0-D8 force breaks the conformal invariance of $R^4$
and so the instanton size is no longer a moduli.
With the 8-0 strings attached to the D0-brane, the linear term in
\QMpot\ is cancelled and so
the moduli space is not lifted, but we
still expect that if the
D0 tries to get fat inside the
D4-brane, conformal symmetry of $R^4$
will be broken and size will not be a modulus of
the instanton.

\subsec{Chiral symmetry breaking and five dimensional Chern-Simons
theory on a boundary.}

Let us put a boundary on the 4+1 dimensional system.
The anomaly equation in 3+1 dimensions
\eqn\anomalyThree{\dd^{0}J_{0}^5 = F_{01}F_{23}}
as in 1+1 dimensions has an interpretation as
coming from a 5d bulk current $J^5$ transverse
to the 3+1 boundary. Again in the 3+1 theory it appears as
though charge in created from no where, but it is really coming
in from the 5th dimension.
Charge can leave the left-hand-side three dimensional wall and enter the
right hand side 3-wall. This is precisely
the Goldstone-Wilczek current \gw.

In the
brane set-up, to add a boundary, we add NS 5-branes so that
the D4-brane can end on them, creating a 3+1 wall.
The exact set-up is
NS 5-branes in directions 012345,
$N_c$ D4-brane in 01236,
$N_f$ D8-branes in 012345689.
In order to satisfy
the massive IIA equations of motion
the NS 5-branes must have D6-branes
ending on them in directions
0123789.
Now 6-4 strings
are the massless 3+1 chiral fields
on the boundary. The
left-side has left-handed chiral superfields
and the right-side has right-handed chiral
superfields \bh. This set-up was discussed in \az.
However, if the 6-4 string representing
a left-handed chiral superfield acquires
some energy it can become a 8-4 string which in
the D4-brane world volume theory looks like
a massive five dimensional
field. Then it can
move to the other side of the
interval and turn back into
a 6-4 string
becoming massless again.
However, now it has become a right-handed
chiral superfield. The superfield
has changed chirality!
But because of this big potential
barrier proportional to the mass of the 5d field
it is a rather unlikely event. It can become much more likely
if we attach the string to a D0-brane. Since
the D0-brane is electrically charged due to the
5d Chern-Simons term it wants to
bind to the oppositely charged string making a
BPS state. The heavy D0-brane can then carry a relatively light string
across the interval. Moreover, the D0-brane has exactly the
right potential to be an instanton effect.
\eqn\inst{V = e^{-{r\over L_6}} = e^{-{1\over g_{YM}^2}}.}
where we have used the relation
between the coupling of the D4-brane $g_{YM}$ to the length of the
interval $L_6$.
We note here that this set-up is very much like
the proposal of \kaplan\ where four-dimensional chiral fermions live on
the boundary of the fifth dimension. The theory in five-dimensions
is a Chern-Simons theory exactly as we have in our brane set-up here.

D0-brane is also similar to a baryon in that it is a composite
of $k$ quarks.

\subsec{Confinement}

There is an interesting mechanism for confinement in 
this picture. Consider massless 4-6 string at one edge of the 
D4-brane and an anti-4-6 string on the other edge.
These represent a quark anti-quark pair. The quarks are separated in the 
5th dimension as well as in $\bf{R^3}$. As $L_6$ becomes of order the 
string scale, it is energetically favorable for the 6-4 string and the 4-6 
string, if they are located in the same place in $\bf{R^3}$, 
to join into a 6-6 string. The 6-6 string is a gauge singlet under 
the color group $SU(N_c)$ but a bifundamental under the 
flavor group $SU(N_f)\times SU(N_f)$. In this 
way, the 6-6 string is a meson.

\newsec{Fun with 8-branes and 2-branes.}

\subsec{D8-D2-D8 bound state}

If we consider the potential for a D2-brane 
in the background of two D8-branes
at positions $+A$ and $-A$, then 
the field theory on the D2-brane will be parity invariant
with two massive fermion fields
\eqn\Pinvar{\cl_{D2} = (X +A)\bar \psi_+\psi_+ 
+ (X - A)\bar\psi_-\psi_-.} 
From the supergravity we 
deduce that 
when one  integrates out the fermions as well as the whole tower 
of string states,
the potential exactly cancels between the 8-branes:
The situation is massless IIA on the inside of the 
walls and massive IIA on the outside.
A D2-brane in the background of a 
flat massless IIA metric has no potential.
The D2-branes sits on the top of a plateau
with steep slopes on either side.
We will see in the next section that 
stringy corrections modify this picture near the 
8-branes. 

\subsec{Meta-stable bound state of D2-D8}

As discussed in
section 3.1 ,
for a D2-brane close to a D8-brane the massive string modes decouple
giving equation \cwPot. This potential actually attracts the D2-brane
to the D8-brane. For the D2-brane far from the D8-brane, stringy modes
couple back in and supergravity tells us that there is
a repulsive force \sugraPot.
Clearly, there is a turn around point and the potential
looks roughly like an upside-down Mexican hat. This allows for a meta
stable D2-brane vacuua that will eventually tunnel to the supersymmetric vacuum at
infinity. This agrees with what was found for the D0-D6 case by
other means \wati.

\newsec{Conclusions}

It would be interesting to pursue the conjectured duality between
the $U(M)$ Non-Abelian Chern-Simons theory at level $k$ and the
fractional quantum Hall effect at level $\nu = {k\over M}$
further. Also refining the picture of non-perturbative chiral
symmetry breaking in the 3+1 theory on the boundary of the 4+1
Chern-Simons theory might be helpful in sorting out how much
chiral symmetry breaking is related to confinement in real QCD.
What is the relationship with the Chern-Simons solitons and
particles we see in 3+1 dimensions? Are they related to baryons in
any way? More speculatively can the fractional quantum Hall effect
be related in any sensible way to chiral symmetry breaking in 3+1
dimensions?

\newsec{Acknowledgements}
I would like to thank O. Aharony, M. Green,
S. Hellerman, A. Karch, B. Kol, M. Schmaltz,
L. Susskind, N. Toumbas, and
S. Thomas for helpful discussions, explanations, and suggestions.
This work was initiated during the 1999 Amsterdam Workshop.
and was supported under DOE grant AE-AC03-76SF-00515.

\listrefs

\end